\documentclass[aps, prl,  showpacs,  superscriptaddress, floatfix,  showkeys,
reprint
]{revtex4-1}

\usepackage{graphicx}
\usepackage{xspace}

\usepackage{amsmath}  
\usepackage{textcomp}  
\usepackage{mathptmx} 
\usepackage[T1]{fontenc}

\graphicspath{{./figures/}}

\begin{document}

\newcommand{\superk}            {Super\nobreakdash-K\xspace}
\newcommand{\neutrino}         {$\nu$\xspace}
\newcommand{\evis}                {$E_{vis}$}
\newcommand{\mutau}             {$\nu_\mu \leftrightarrow \nu_{\tau}$\xspace}
\newcommand{\musterile}        {$\nu_\mu \leftrightarrow \nu_{sterile}$\xspace}
\newcommand{\nue}                 {$\nu_{e}$\xspace}
\newcommand{\numu}              {$\nu_{\mu}$\xspace}
\newcommand{\nutau}              {$\nu_{\tau}$\xspace}
\newcommand{\numutau}         {$\nu_{\mu} \rightarrow \nu_{\tau}$\xspace}
\newcommand{\nubar}              {$\overline{\nu}$\xspace}
\newcommand{\nuebar}            {$\overline{\nu}_{e}$\xspace}
\newcommand{\numubar}         {$\overline{\nu}_{\mu}$\xspace}
\newcommand{\nutaubar}         {$\overline{\nu}_{\tau}$\xspace} 
\newcommand{\tonethree}        {$\theta_{13}$\xspace}
\newcommand{\tonetwo}          {$\theta_{12}$\xspace}
\newcommand{\ttwothree}        {$\theta_{23}$\xspace}
\newcommand{\msqonetwo}     {$\Delta m^2_{21}$\xspace}
\newcommand{\msqonethree}   {$\Delta m^2_{31}$\xspace}
\newcommand{\msqtwothree}   {$\Delta m^2_{32}$\xspace}

\title{Evidence for the Appearance of Atmospheric Tau Neutrinos in Super-Kamiokande}
\date{\today}
\newcommand{\AFFicrr}{\affiliation{Kamioka Observatory, Institute for Cosmic Ray Research, University of Tokyo, Kamioka, Gifu 506-1205, Japan}}
\newcommand{\AFFkashiwa}{\affiliation{Research Center for Cosmic Neutrinos, Institute for Cosmic Ray Research, University of Tokyo, Kashiwa, Chiba 277-8582, Japan}}
\newcommand{\AFFbu}{\affiliation{Department of Physics, Boston University, Boston, MA 02215, USA}}
\newcommand{\AFFbnl}{\affiliation{Physics Department, Brookhaven National Laboratory, Upton, NY 11973, USA}}
\newcommand{\AFFucd}{\affiliation{Department of Physics, University of California, Davis, Davis, CA 95616, USA}}
\newcommand{\AFFuci}{\affiliation{Department of Physics and Astronomy, University of California, Irvine, Irvine, CA 92697-4575, USA }}
\newcommand{\AFFcsu}{\affiliation{Department of Physics, California State University, Dominguez Hills, Carson, CA 90747, USA}}
\newcommand{\AFFcnm}{\affiliation{Department of Physics, Chonnam National University, Kwangju 500-757, Korea}}
\newcommand{\AFFduke}{\affiliation{Department of Physics, Duke University, Durham NC 27708, USA}}
\newcommand{\AFFfukuoka}{\affiliation{Junior College, Fukuoka Institute of Technology, Fukuoka, 811-0214, Japan}}
\newcommand{\AFFgmu}{\affiliation{Department of Physics, George Mason University, Fairfax, VA 22030, USA }}
\newcommand{\AFFgifu}{\affiliation{Information and Multimedia Center, Gifu University, Gifu, Gifu 501-1193, Japan}}
\newcommand{\AFFuh}{\affiliation{Department of Physics and Astronomy, University of Hawaii, Honolulu, HI 96822, USA}}
\newcommand{\AFFkanagawa}{\affiliation{Physics Division, Department of Engineering, Kanagawa University, Kanagawa, Yokohama 221-8686, Japan}}
\newcommand{\AFFkek}{\affiliation{High Energy Accelerator Research Organization (KEK), Tsukuba, Ibaraki 305-0801, Japan }}
\newcommand{\AFFkobe}{\affiliation{Department of Physics, Kobe University, Kobe, Hyogo 657-8501, Japan}}
\newcommand{\AFFkyoto}{\affiliation{Department of Physics, Kyoto University, Kyoto, Kyoto 606-8502, Japan}}
\newcommand{\AFFumd}{\affiliation{Department of Physics, University of Maryland, College Park, MD 20742, USA }}
\newcommand{\AFFmit}{\affiliation{Department of Physics, Massachusetts Institute of Technology, Cambridge, MA 02139, USA}}
\newcommand{\AFFmiyagi}{\affiliation{Department of Physics, Miyagi University of Education, Sendai, Miyagi 980-0845, Japan}}
\newcommand{\AFFnagoya}{\affiliation{Solar Terrestrial Environment
Laboratory, Nagoya University, Nagoya, Aichi 464-8602, Japan}}
\newcommand{\AFFnagoyaKM}{\affiliation{Kobayashi-Maskawa Institute for the Origin of Particle 
and the Universe,  Nagoya University, Nagoya, Aichi 464-8602, Japan}}
\newcommand{\AFFsuny}{\affiliation{Department of Physics and Astronomy, State University of New York, Stony Brook, NY 11794-3800, USA}}
\newcommand{\AFFniigata}{\affiliation{Department of Physics, Niigata University, Niigata, Niigata 950-2181, Japan }}
\newcommand{\AFFokayama}{\affiliation{Department of Physics, Okayama University, Okayama, Okayama 700-8530, Japan }}
\newcommand{\AFFosaka}{\affiliation{Department of Physics, Osaka University, Toyonaka, Osaka 560-0043, Japan}}
\newcommand{\AFFseoul}{\affiliation{Department of Physics, Seoul National University, Seoul 151-742, Korea}}
\newcommand{\AFFshizuokasc}{\affiliation{Department of Informatics in
Social Welfare, Shizuoka University of Welfare, Yaizu, Shizuoka, 425-8611, Japan}}
\newcommand{\AFFskk}{\affiliation{Department of Physics, Sungkyunkwan University, Suwon 440-746, Korea}}
\newcommand{\AFFtohoku}{\affiliation{Research Center for Neutrino Science, Tohoku University, Sendai, Miyagi 980-8578, Japan}}
\newcommand{\AFFtokyo}{\affiliation{The University of Tokyo, Bunkyo, Tokyo 113-0033, Japan }}
\newcommand{\AFFipmu}{\affiliation{Kavli 
Institute for the Physics and Mathematics of the Universe 
(WPI), 
University of Tokyo, Kashiwa, Chiba, 277-8583, Japan}}
\newcommand{\AFFtokai}{\affiliation{Department of Physics, Tokai University, Hiratsuka, Kanagawa 259-1292, Japan}}
\newcommand{\AFFtit}{\affiliation{Department of Physics, Tokyo Institute
for Technology, Meguro, Tokyo 152-8551, Japan }}
\newcommand{\AFFtsinghua}{\affiliation{Department of Engineering Physics, Tsinghua University, Beijing, 100084, China}}
\newcommand{\AFFwarsaw}{\affiliation{Institute of Experimental Physics, Warsaw University, 00-681 Warsaw, Poland }}
\newcommand{\AFFNCNR}{\affiliation{National Centre For Nuclear Research, 00-681 Warsaw, Poland}}
\newcommand{\AFFuw}{\affiliation{Department of Physics, University of Washington, Seattle, WA 98195-1560, USA}}
\newcommand{\AFFuam}{\affiliation{Department of Theoretical Physics, University Autonoma Madrid, 28049 Madrid, Spain }}

\AFFicrr
\AFFkashiwa
\AFFuam
\AFFbu
\AFFbnl
\AFFuci
\AFFcsu
\AFFcnm
\AFFduke
\AFFfukuoka
\AFFgifu
\AFFuh
\AFFkek
\AFFkobe
\AFFkyoto
\AFFmiyagi
\AFFnagoya
\AFFnagoyaKM
\AFFsuny
\AFFokayama
\AFFosaka
\AFFseoul
\AFFshizuokasc
\AFFskk
\AFFtokai
\AFFtokyo
\AFFipmu
\AFFtsinghua
\AFFwarsaw
\AFFNCNR
\AFFuw
%

\author{K.~Abe}
\author{Y.~Hayato}
\AFFicrr
\AFFipmu
\author{T.~Iida}
\author{K.~Iyogi} 
\AFFicrr
\author{J.~Kameda}
\author{Y.~Koshio}
\AFFicrr
\AFFipmu
\author{Y.~Kozuma} 
\author{Ll.~Marti} 
\AFFicrr
\author{M.~Miura} 
\author{S.~Moriyama} 
\author{M.~Nakahata} 
\author{S.~Nakayama} 
\author{Y.~Obayashi} 
\author{H.~Sekiya} 
\author{M.~Shiozawa} 
\author{Y.~Suzuki} 
\author{A.~Takeda} 
\AFFicrr
\AFFipmu
% ...

\collaboration{The Super-Kamiokande Collaboration}
\noaffiliation

\begin{abstract}
  Super-Kamiokande atmospheric neutrino data were fit with an
  unbinned maximum likelihood method to search for the appearance of
  tau leptons resulting from the interactions of oscillation-generated
  tau neutrinos in the detector. Relative to the expectation of unity,
  the tau normalization is found to be $1.42 \pm 0.35 \ (stat) {\
  }^{+0.14}_{-0.12}\ (syst) $ excluding the no-tau-appearance
  hypothesis, for which the normalization would be zero, at the
  3.8$\sigma$ level.  We estimate that $180.1 \pm 44.3\ (stat) {\
  }^{+17.8}_{-15.2}\ (syst)$ tau leptons were produced in the 22.5~kton
  fiducial volume of the detector by tau neutrinos during the 
  2806 day running period.
  In future analyses, this large sample of selected tau events will
  allow the study of charged current tau neutrino interaction physics
  with oscillation produced tau neutrinos.
\end{abstract}

\pacs{14.60.Pq, 96.40.Tv} 

\keywords{Neutrino Oscillations, Super-Kamiokande, Tau Neutrinos,
  Atmospheric Neutrinos}

\maketitle

It is now well known that neutrinos undergo flavor oscillations. The
flavor states of the neutrino measured through the weak interaction
are quantum mechanical mixtures of neutrino mass states.  As observed
in the quark sector, this mixture results in oscillations of detected
flavor states.  Evidence exists for this effect in atmospheric
neutrinos~\cite{Fukuda:1998mi,Ashie:2005ik}, solar
neutrinos~\cite{Hosaka:2005um,
  Abe:2010hy,Ahmad:2002jz,Aharmim:2009gd,Aharmim:2011vm}, reactor
experiments~\cite{:2008ee}, and long-baseline oscillation
experiments~\cite{Ahn:2006zza, Adamson:2008zt,Abe:2012gx}.  In 2011,
the T2K~\cite{Abe:2011sj}, MINOS~\cite{Adamson:2011qu}, and Double
Chooz~\cite{Abe:2011fz} experiments showed the first indications of
full three-flavor oscillations.
In 2012 the Daya Bay~\cite{An:2012eh} and RENO~\cite{Ahn:2012nd}
experiments reported the first precision measurements of the
$\theta_{13}$ mixing angle which drives three-flavor oscillation.

Definitive proof of flavor oscillation requires unambiguous
appearance of the charged current interaction of a neutrino not in the
original source.
In the dominant oscillation for \numu at GeV energies, \numutau
oscillations, observing the resulting $\tau$ lepton is quite
difficult.  This is because producing a tau lepton requires a neutrino
of energy greater than a threshold of 3.5 GeV.
Long-baseline experiments tuned to the neutrino oscillation maximum
for their distances tend to have the bulk of their neutrinos below
this energy. Furthermore, the tau lepton immediately decays to final
states with an electron, muon or mesons plus a tau neutrino so the tau
lepton itself cannot be easily seen.
Nevertheless, the OPERA Collaboration was recently able to show
evidence for a single reconstructed event in their emulsion consistent
with tau appearance~\cite{Collaboration:2011ph}.
%
The Super-Kamiokande (Super-K) Collaboration first published a search for tau
appearance in atmospheric neutrinos in 2006~\cite{Abe:2006fu}.  Since
the atmospheric neutrino flux extends to energies well above 10~GeV,
and spans a wide range of baselines, we expect to see tau leptons
produced in the Super-K detector.  However, these events must be
distinguished from other high-energy atmospheric neutrino
interactions. Further comparisons of these techniques can be found
in~\cite{Migliozzi:2011bj} and prospects for future detectors can be
found in~\cite{Conrad:2010mh}.

This Letter reports a result from a new search utilizing the
Super-Kamiokande experiment.  
This analysis addresses the question of
whether the atmospheric data are consistent with the lack of
oscillation-generated \nutau or whether they are necessary to
explain the observations.
\superk is a 50\,000
ton water Cherenkov detector\cite{Fukuda:2002uc} with 22.5~kton of
fiducial volume.  It consists of two concentric detectors:  an
inner detector with 11\,129 inward-looking 20 in.
photodetectors and an outer detector with 1885
outward-facing 8 in. photodetectors which acts as a veto.  Its
large target mass makes it well suited to look for the rare appearance
of tau neutrinos from oscillations.
The typical energy of atmospheric neutrinos is about 1~GeV.  Because of
the previously noted energy threshold,
about one \nutau
charged current event per kton-yr should be
produced in the \superk detector.


\superk has been in operation for approximately 15 years and has
had several running configurations indicated by the labeling SK-I
(1996-2001), SK-II (2002-2005), SK-III (2006-2008) and SK-IV
(2008-2012).  The previously reported \superk
result~\cite{Abe:2006fu}
was based on the data from
SK-I alone.  Since that time the analysis has been improved to
increase its sensitivity and the data set has been expanded to
also include SK-II and SK-III, thereby almost doubling its size.  As
the total data set covers the period between 1996 and 2008, it
comprises 2806 days of live time.


In order to predict the rate of both the tau signal and atmospheric
background, a full Monte Carlo (MC) simulation is used both to predict
the neutrino interactions inside the detector and to model the
response of \superk itself.  Three-dimensional neutrino fluxes for
\numu and \nue produced in atmospheric showers are taken from the flux
calculation of Honda {\it et al.}~\cite{Honda:2004yz}.  The fluxes are
oscillated with a custom code~\cite{prob3++} which takes into account
all relevant path lengths, energies and matter effects using our
current knowledge of three-flavor neutrino oscillation parameters.
The oscillation parameters used are~\cite{Wendell:2010md,
  Abe:2011ph,Schwetz:2011zk} \msqtwothree $ \rm = 2.1 \times 10^{-3}
eV^2$, \msqonetwo $ \rm = 7.6 \times 10^{-5} eV^2$, $\rm \sin^2 2
\theta_{23} = 1.0$, $\rm \sin^2 2 \theta_{12 } = 0.84$, $\rm
\delta_{CP} = 0$.
The Super-K best fit value of \msqtwothree from
    \cite{Wendell:2010md, Abe:2011ph} was used in order to make use
  of the full set of systematic errors which were previously evaluated
  around this point.  However, the difference in results between using
  this value and that of recent more precise values reported in the
  literature~\cite{Adamson:2011ig} is found to be negligible
  due to the wide range of L and E sampled by the atmospheric
  neutrinos.
For the values of
\tonethree, recent Daya Bay~\cite{An:2012eh} and
RENO~\cite{Ahn:2012nd} results are combined in a weighted average and
we use $\sin^2 2 \theta_{13} = 0.099$.
The interactions of the \numu, \nue, and oscillation-produced {\nutau}s
with the nuclei of water molecules inside the \superk detector are
modeled with the NEUT~\cite{Ashie:2005ik, Hayato:2002sd} neutrino
interaction code.  Finally, a GEANT3~\cite{Brun:1987ma} based detector
MC is used to simulate \superk itself.  More detailed
descriptions of this software can be found in~\cite{Ashie:2005ik}.

For the purposes of this analysis it is important to understand some
details of the neutrino interaction model.  The NEUT code models the
known neutrino-nucleon interactions including quasielastic
scattering, single meson production, coherent pion production, and
deep-inelastic scattering (DIS).  All \numu and \nue interactions are
simulated.  Additionally, charged-current (CC) \nutau
interactions are simulated and added to the sample using weighting
based on the oscillation probabilities.  Neutral current (NC)
interactions are assumed to be unaffected by oscillations.
The \nutau CC cross~sections are calculated following the same models
as those used for \numu and \nue with the appropriate lepton mass
terms. In the case of single and coherent pion production lepton mass
corrections not included in the original models are also
employed~\cite{Berger:2007rq,Rein:2006di}.
Tau leptons are decayed using  TAUOLA (Version
2.6)~\cite{Jadach:1993hs}. Since the distribution
of decay particles depends on the polarization of the tau lepton,
a polarization model from \cite{Hagiwara:2003di} is incorporated into NEUT.
At the relevant neutrino energies selected by
this analysis, the CC interactions contain a high percentage of DIS
(46\%) 
with the portion of CC events induced by the \nutau 
signal interactions alone containing 56\% DIS.  In
the calculation of the cross sections of DIS, the
GRV94~\cite{Gluck:1994uf} parton distribution functions are used, and
additional corrections to make the DIS cross sections match smoothly
with the resonance region as developed by Bodek and Yang are also
applied~\cite{Bodek:2002vp}.  More details of the DIS implementation
can be found in~\cite{Ashie:2005ik}.


The signature of oscillation-induced tau neutrinos in the atmospheric
flux is the detection of the decay of tau leptons in the \superk
detector.  As the leptonic decays of the tau look on the whole very
similar to normal CC DIS interactions from \numu and \nue, an analysis
is developed which attempts to select the hadronic decays. 

In order to select the tau events, we first 
identify high energy events contained in the inner detector by
requiring
that there is no appreciable activity
in the outer detector, the interaction is in the fiducial volume
(the distance to the nearest wall > 200~cm), and  the event has
more than 1.3 GeV of visible energy.  The selection efficiencies for
this set of cuts are 81\% for the \nutau CC signal and 23\% for the
background events respectively.

The presence of the extra pions from hadronic tau decay which come
from a heavy object can statistically separate the signal from the
normal \numu and \nue  CC and NC background.
In order to further separate the signal from the background, a set
of variables which show differentiation between the two samples is
used as the inputs to a feed-forward neural network (NN).  
The NN is configured using the TMVA
package~\cite{Hocker:2007ht} with seven inputs nodes, one hidden layer
with 10 nodes and one output node.  Exclusive training and testing
samples are selected from the MC sets to avoid bias and test
for overtraining.

The variables used are 
(a) the log base 10 of the total visible energy of the event, 
(b) the particle identification of the maximum energy ring in the event,
(c) the number of decay electron candidates in the event,
(d) the maximum distance between the primary interaction and any decay
electron found from a pion or muon decay, 
(e) the clustered sphericity of the event in the center of mass system, 
(f) the number of possible Cherenkov ring fragments, 
and finally
(g) the fraction of total number of photoelectrons in the events carried by the first ring.   
The agreement between downward going data and MC simulations (where no
tau signal is expected) for the NN output along with the overlaid
expected tau signal is shown in Fig.~\ref{fig:nnoutput}.  See
Supplemental Material [Fig.~\ref{fig:variables} at the end of this
Letter] for the additional agreement between the data and MC
simulations of the seven input variables to the NN.

\begin{figure}[!hbp]
  \includegraphics[width=2.8in]{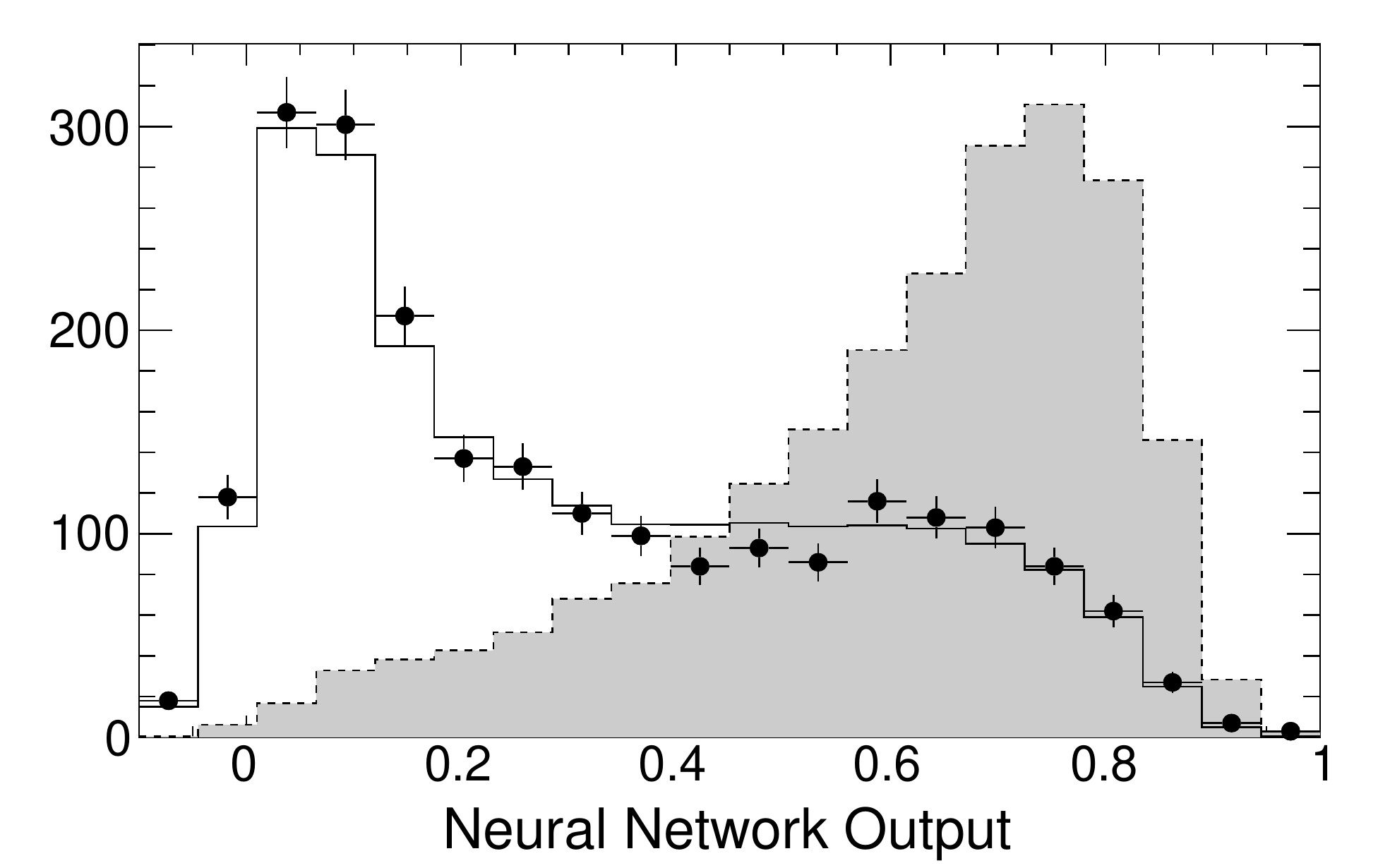}
  \caption{The separation of signal and background by the neural
    network. The downward going data (points) are overlaid with the
    downward going atmospheric MC simulations (solid line). Also shown is the tau
    signal MC simulation (shaded). The tau signal is normalized for equal
    statistics.}
  \label{fig:nnoutput}
\end{figure}


All of the oscillation-induced tau neutrinos will come from below due
to differing path lengths in the earth.  In order avoid encoding such
up-down biases into the network, and to select events based
solely on their topology, the training is performed by weighting the
oscillation probabilities of all events based on their energies alone,
not their direction.
In this way, all oscillation probabilities are correct on average, but
upward-going and downward-going events are treated the same in the
training process.  This technique has the added benefit of not setting
the weights of the down-going signal events to zero, thus preserving
MC statistics. The training is performed such that a NN
output which is near 1.0 signifies that the event is taulike, while
events near 0.0 are nontaulike. After training, the NN is found to
efficiently separate the tau appearance signal from the background of
other atmospheric neutrino interactions.

When acting on the events passing the preselection cuts, 75\% of the
signal events (60\% total efficiency) and only 26\% of the background
events (6\% total efficiency) remain when events with a NN output of
greater than 0.5 are considered. In this ``taulike'' sample, NC
background makes up 26\% of the sample and is an important remaining
background.  Table~\ref{table:breakdown} further displays the
fractional breakdown of the interaction modes in the sample. In order
to extract maximum information from the event samples, instead of
cutting on the NN output, the output of the NN is combined with the
zenith direction of the event into a probability distribution
functions (PDF) and is used to jointly fit the tau and background
components.

\begin{table}[!htbp]
   \begin{tabular}{lrrr}
     \hline 
     \hline
     Interaction mode  &  {\hskip 0.25in} NN < 0.5  & {\hskip 0.25in} NN > 0.5 & {\hskip 0.25in} All\\
     \hline
     CC \nue      &    781.4 (0.40)    &  381.3  (0.46)   & 1162.7  (0.42)\\ 
     CC \numu   &  1070.2 (0.55)    &  200.2  (0.24)   & 1270.4  (0.46)\\
     CC \nutau   &     12.4  (0.01)    &    37.2  (0.04)   &    49.7   (0.02)\\
     NC              &      95.2 (0.05)    &  209.3   (0.25)   &  304.4  (0.11)\\
     \hline
     \hline
   \end{tabular}
   \caption{The fractional breakdown of interaction modes of both the expected
     signal (CC \nutau) and background for the SK-I period.  For fitting purposes the entire sample is
     used in the analysis, but the NN enhanced (NN >0.5) and depleted  (NN<0.5) signal
     selections are shown here to demonstrate the effect of signal and background
     separation.  For each sample, the number of selected SK-I MC events is shown scaled by the
     1489 days of SK-I live time.  The fractional breakdown by interaction mode of
     each sample is shown in parentheses.}
   \label{table:breakdown}
\end{table}

\begin{figure}[!t]
  \includegraphics[width=3.5in]{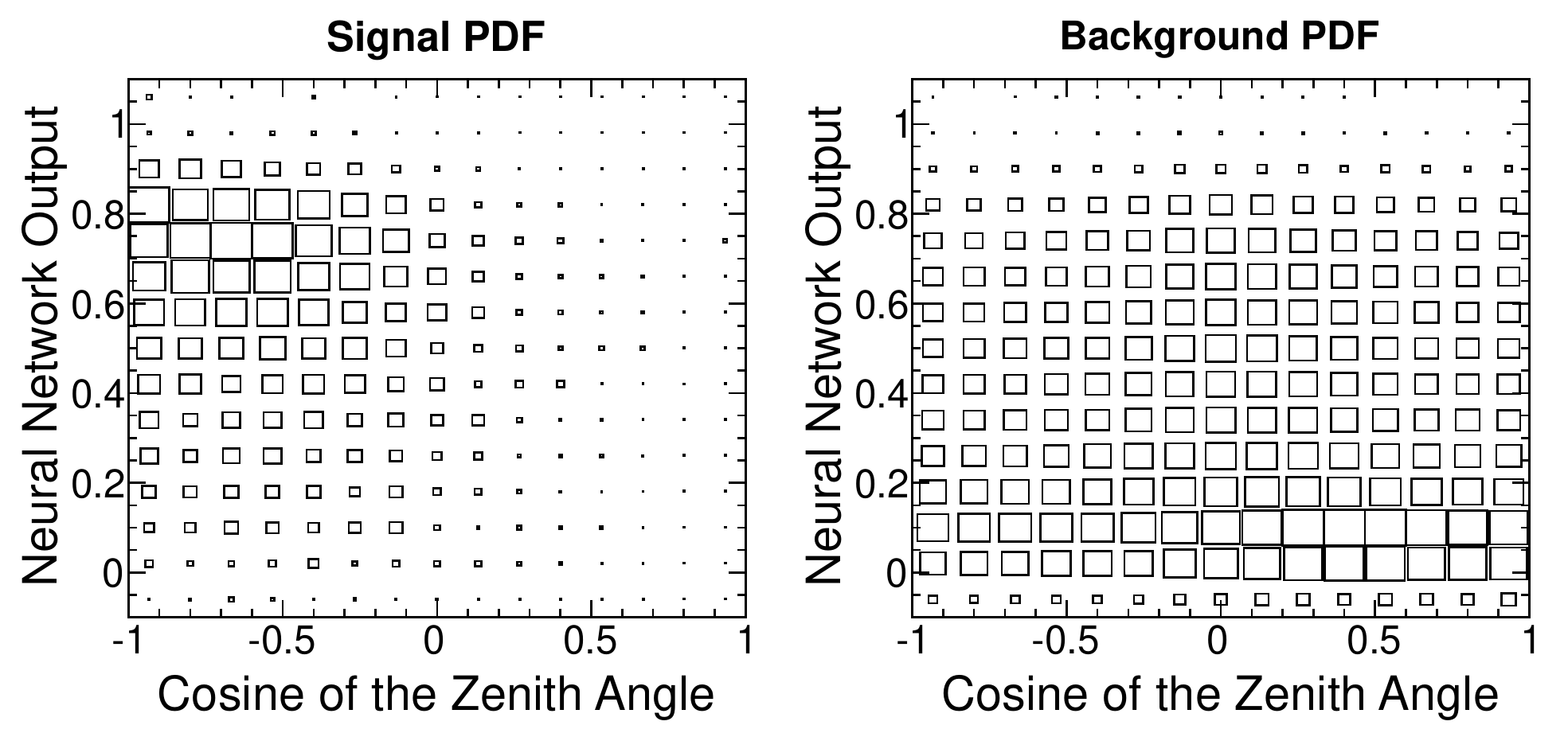}  
  \caption{Histograms of the PDFs of both tau signal (left) and
    atmospheric background (right).  The vertical axis is the output
    of the NN, the horizontal axis the cosine of the event
    zenith direction. Upward going events are to the left, downward
    going to the right. The tau signal appears in the upward-going
    tau-like region.}
  \label{fig:signal-vs-background}
\end{figure}


An example of the two-dimensional distributions of the NN output
versus the direction of the detected events used to discriminate
signal from background is shown in
Fig.~\ref{fig:signal-vs-background}.  Distributions for both
oscillation-generated taus on the left  and other atmospheric
background on the right are shown. The vertical axes of these
two-dimensional distributions contain the output of the NN and
reflects how taulike the event is (NN output near 1.0 taulike,
0.0 nontaulike).  The horizontal axis is the cosine of the
reconstructed zenith angle of the event which is determined by an
energy-weighted sum of the ring directions in the event.
The tau events (left-hand panel) are indicated as taulike by the NN and
come from below [cos($\theta$) near -1.0] as expected.  
In contrast, other atmospheric neutrinos (right-hand panel)
are primarily nontaulike
and come from both above and below.  In fact, it can be seen that
these events are depleted in the upward-going direction due to their
oscillation into (mostly noninteracting) tau neutrinos.
By varying the relative normalization of the two distributions both
the amount of tau appearance and the overall background level can be
adjusted.

PDFs for each run period for both signal and background are built out
of two-dimensional histograms prepared from the MC simulations, with the
probability density following the normalized bin contents.  Then, an
unbinned likelihood fit of the data is done to the sum of the signal
and background PDFs varying the normalization between them.  It is
necessary to perform an unbinned fit as statistics of bins in the
full two-dimensional space would be quite low.  The result of the fit
is a normalization factor on the signal and the background which tells
us how many tau interactions are needed to be consistent with our data
set.  Separate PDFs are produced for SK-I, -II, and -III, and each data
set is fit to its appropriate MC set.  The data sets are fit both
individually for each run period and jointly together.

Although the technique employed here is more sophisticated than that
of~\cite{Abe:2006fu}, it is also more sensitive to some systematic
errors since large numbers of background events remain in the
nonsignal regions of the fitting space which were previously removed
by cuts.
By training the NN to recognize tau interactions, the
NN also learns to effectively separate quasielastic
from multi-pi and DIS interactions
in the background samples.
This is because the DIS events
tend to have many extra pions in them, and thus look more like the
tau signal.
The DIS portion of the interactions thus forms a large part of the
background in the signal region and we therefore explicitly take into
account uncertainties in the DIS normalization in the fit.

The average neutrino energy in the DIS interactions in our sample is 14 GeV and
the cross section is not known to better than the 10\% level at that
energy.  We also know that the application of the Bodek-Yang
corrections~\cite{Bodek:2002vp} tends to suppress our DIS interactions
at higher energies by about 5\%.  For this reason, the DIS error is
introduced into the fit as a 10\% Gaussian error constraint.  After
the fit is completed it is found that the amount of DIS is increased
from its nominal value by 10\% at the best fit point.  If the fit is
performed with no constraint on the DIS fraction at all, then the DIS
fraction fits 14\% higher than the nominal value.


The fit is performed on each data period separately, and is also
performed jointly with all data periods being fit at the same time.
In the case of finding the
exact normalization as predicted by the MC simulations, these factors would
be 1.0.  When the data periods are fit together, the tau normalization
is found to be $1.42 \pm 0.35\ (stat)$ with the background
normalization  $0.94 \pm 0.02\ (stat)$.  When fit separately, the tau
normalizations are found to be    $1.27 \pm 0.49\ (stat)$, $1.47 \pm 0.62\ (stat)$,
and $2.16 \pm 0.78\ (stat)$ for SK-I, SK-II and SK-III respectively.

\begin{figure}[!htbp]
  \includegraphics[width=3.1in]{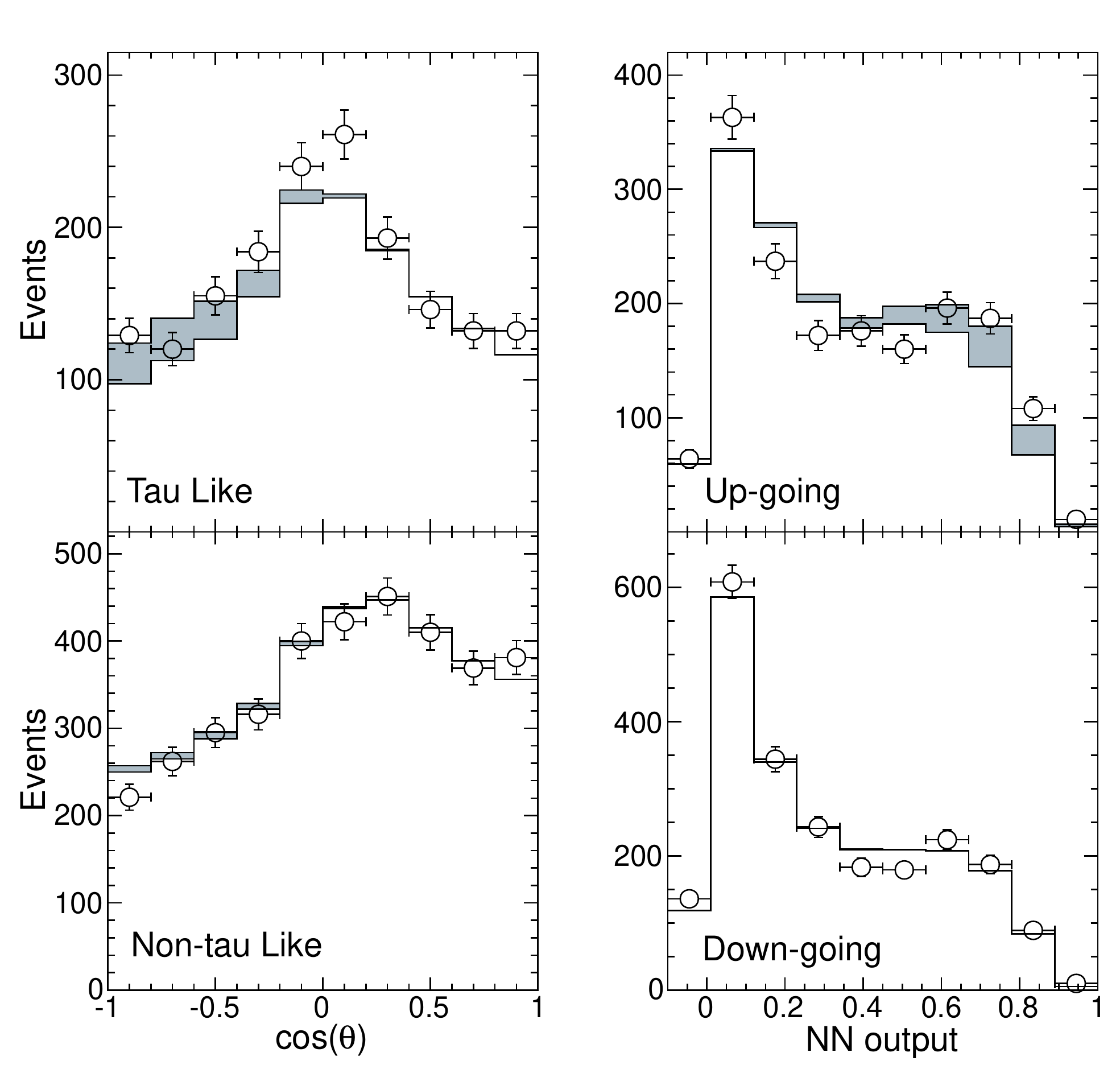}
  \caption{Fit results showing projections in the NN output and zenith
    angle distribution for taulike (NN>0.5), upward-going
    [$\cos(\theta) < -0.2$], nontaulike (NN<0.5), and downward-going
    [$\cos(\theta) > 0.2$] events for both the two-dimensional PDFs
    and data.  The PDFs and data sets have been combined from SK-I
    through SK-III in this figure.  The fitted tau signal is shown in
    gray.}
  \label{fig:4-panel}
\end{figure}

It is also instructive to examine the results of the combined fit graphically.
Binned projections of the fitted results can illustrate the quality
and features of the fit.  Figure~\ref{fig:4-panel} shows the
projections in zenith for both taulike (NN output > 0.5) and
nontaulike (NN output < 0.5) events, along with the projections in
NN output for both upward-going [$\cos(\theta) < -0.2$] and
downward-going [$\cos(\theta) > 0.2$] events.  In these plots the PDFs
have been rescaled to the fitted normalization values.  The fitted tau
signal is shown in gray.  Good agreement is seen in all
distributions. As expected, tau events are observed as an excess of
taulike events in the upward-going direction.
In these plots the PDFs and data sets
from all of the run periods have been combined together.
 
There are 28 uncertainties which are a subset of those used in the
\superk three-flavor atmospheric neutrino analysis. A more detailed
description of them can be found in~\cite{Wendell:2010md}.
The systematic errors for the analysis are divided into two sets.  The
first set, which describes errors on the tau expectation itself, plays
no role in comparing the fitted observed number of events with the
no-tau-appearance hypothesis and does not affect the significance of this
quoted result.
However, this set is used to quote an error on the
expected number of events and includes uncertainties in the \nutau cross
section  and any uncertainty that would increase both the signal and
the background in a way that does not change the significance of the
reported result.
Detector biases on selection and fitting are included in these uncertainties
but are quite small compared to the tau cross-section error, the
largest being a 5\% error on the detector energy scale.
The error on the tau cross section was made by a comparison of
NEUT~\cite{Hayato:2002sd} with several other models, looking in
particular at the differences between NEUT and the cross-section model
by Hagiwara {\it et al}~\cite{Hagiwara:2003di}. Another comparison
between cross-section models was recently completed by the authors
of~\cite{Conrad:2010mh} and gave similar results.  
As noted above, this 25\% error does not contribute to the
reported significance of this Letter. However, future analysis using
this high statistics data set employing full simultaneous
treatment of all relevant systematic errors can measure this
cross section and constrain its uncertainty using the \superk
data itself.

The second class of errors includes those that would affect the observed
signal but not the background, or otherwise would cause the
significance of the measured normalization to change when doing the
fit. 
There are five such errors, all expressed as ratios: upward to
downward neutrino flux, horizontal to vertical neutrino flux, kaon to
pion originated neutrino flux, NC to CC cross section, and the upward
to downward detector energy scale difference.
In the current analysis the dominant error on the signal was the NC/CC
ratio changing the fitted number of events of about $\pm 7\%$ due to
the relatively large percentage of NC background in the signal region.

Also included in the errors which can change the measured results and
significance are those due to variations in the known oscillation
parameters.  For this study they are varied within the 1$\sigma$
limits of a combined SK-I+SK-II+SK-III atmospheric oscillation
analysis result assuming the normal hierarchy~\cite{Abe:2011ph}.  The
\msqtwothree is varied between $1.92 \times 10^{-3}$ and $2.22 \times
10^{-3} {\rm eV}^2$, $\sin^2 2 \theta_{23}$ is varied between $0.93$
and $1.0$.  The \tonethree values are varied within our combined Daya
Bay~\cite{An:2012eh} and RENO~\cite{Ahn:2012nd} results of $\sin^2 2
\theta_{13} = 0.099 \pm .014$. The use of nonzero \tonethree results
in a 13\% reduction of the fitted normalization as three-flavor
oscillations produce high energy upward-going electron neutrinos which
add to the upward-going background, thus decreasing the needed number
of tau neutrinos to explain the signal region.  However, the variation
in \tonethree around this central value results in less than a 1\%
change in the fit result. For this analysis, we set the value of
$\delta_{CP}$ to zero.  Varying the value of $\delta_{CP}$ results in,
at most, a 1.3\% difference in the number of fitted taus, and we
neglect this uncertainty.  The systematic errors are summarized in
Table~\ref{table:systematics}.

\begin{table}[htbp]
    \begin{tabular}{lrr}
      \hline 
      \hline
      Systematics uncertainties for \nutau normalization       & ~~$+$ \% & ~~$-$ \%\\
      \hline
      \superk atmospheric $\nu$ oscillation errors &   &  \\
      {\hspace{0.5cm}} 28 error terms {\hspace{1.6cm}} (expected  events)  & 13.4 & 14.7\\ 
      {\hspace{0.5cm}} 5 error terms  {\hspace{1.75cm}} (observed events) &   7.9  & 8.5\\ 
      \hline
      Tau neutrino cross section {\hspace{0.675cm}} (expected events)      & 25.0 & 25.0\\
      \hline
      Oscillation parameters {\hspace{1.1cm}} (observed events)   &      5.4   & 1.3 \\
      \hline 
      \hline
    \end{tabular}
    \caption{Summary of systematic uncertainties for both the expected
      and observed number of \nutau events. The errors for each
      category including that of the oscillation parameters have been added in
      quadrature. }
    \label{table:systematics}
\end{table}

Including and combining the observed (+9.6 -8.6\% ) and expected
(+28.4 -30.0\%) systematic uncertainties separately, the fitted value
of the tau normalization is $ 1.42 \pm 0.35\ (stat) {\
}^{+0.14}_{-0.12} \ (syst)$.  After rescaling the MC by all
fitting factors 
and correcting for efficiency, the observed number of fitted events
over the entire running period is calculated to be $180.1 \pm 44.3\
(stat) {\ }^{+17.8}_{-15.2}\ (syst)$ events.  This is to be compared to
an expectation of $120.2{\ }^{+34.2}_{-34.8}\ (syst)$ interactions in
the fiducial volume if no fitting factors are applied.
Identifying this large statistics sample opens the possibility to
  study charged current tau neutrino interaction physics with
  oscillation produced tau neutrinos.
%


The observed number of events is converted to the
significance level at which we can reject the no-tau-appearance
hypothesis.  The measured signal normalization (1.42 and its
associated statistical and systematic errors) is compared with the
case of no \nutau appearance, which would have a normalization of zero.  
An asymmetric Gaussian centered at 1.42 is prepared and the integral
of the PDF below zero is calculated.
The $p$~value is $6.2 \times 10^{-5}$ which
corresponds to a significance level of 3.8$\sigma$.  
A significance of 2.7$\sigma$ is expected for the nominal expected
signal. The larger measured significance is a consequence of the fact
that more signal was measured than expected.
The DIS fraction is fit with a 10\% increase over its nominal value,
and is correlated with the tau normalization.  Because of this, not
only is the fitted tau normalization lower than it would be without
this error, but the error on the tau normalization is larger than it
would otherwise be due to the presence of the correlated DIS error,
thus slightly reducing the measured significance.  It should be noted
that if the inverted hierarchy is chosen instead of the normal one
when calculating the oscillation probabilities, the expected number of
\tonethree-induced upward-going electrons is reduced, approximately in
half, resulting in a somewhat higher fitted value (1.56) and a
correspondingly higher significance.


In summary, we find that the Super-Kamiokande atmospheric neutrino
data are best described by neutrino oscillations that include tau
neutrino appearance in addition to the overwhelming signature of muon
neutrino disappearance.  By a neural network analysis on the zenith
angle distribution of multi-GeV contained events, we have demonstrated
this at a significance of  3.8$\sigma$.

\begin{acknowledgments} 
  We gratefully acknowledge the cooperation of the Kamioka Mining and
  Smelting Company.  The Super-Kamiokande experiment has been built
  and operated from funding by the Japanese Ministry of Education,
  Culture, Sports, Science and Technology, the U.S.
  Department of Energy, and the U.S. National Science Foundation.
\end{acknowledgments}

\bibliography{references}

\begin{figure}[!htbp]
  \includegraphics[width=4.0in]{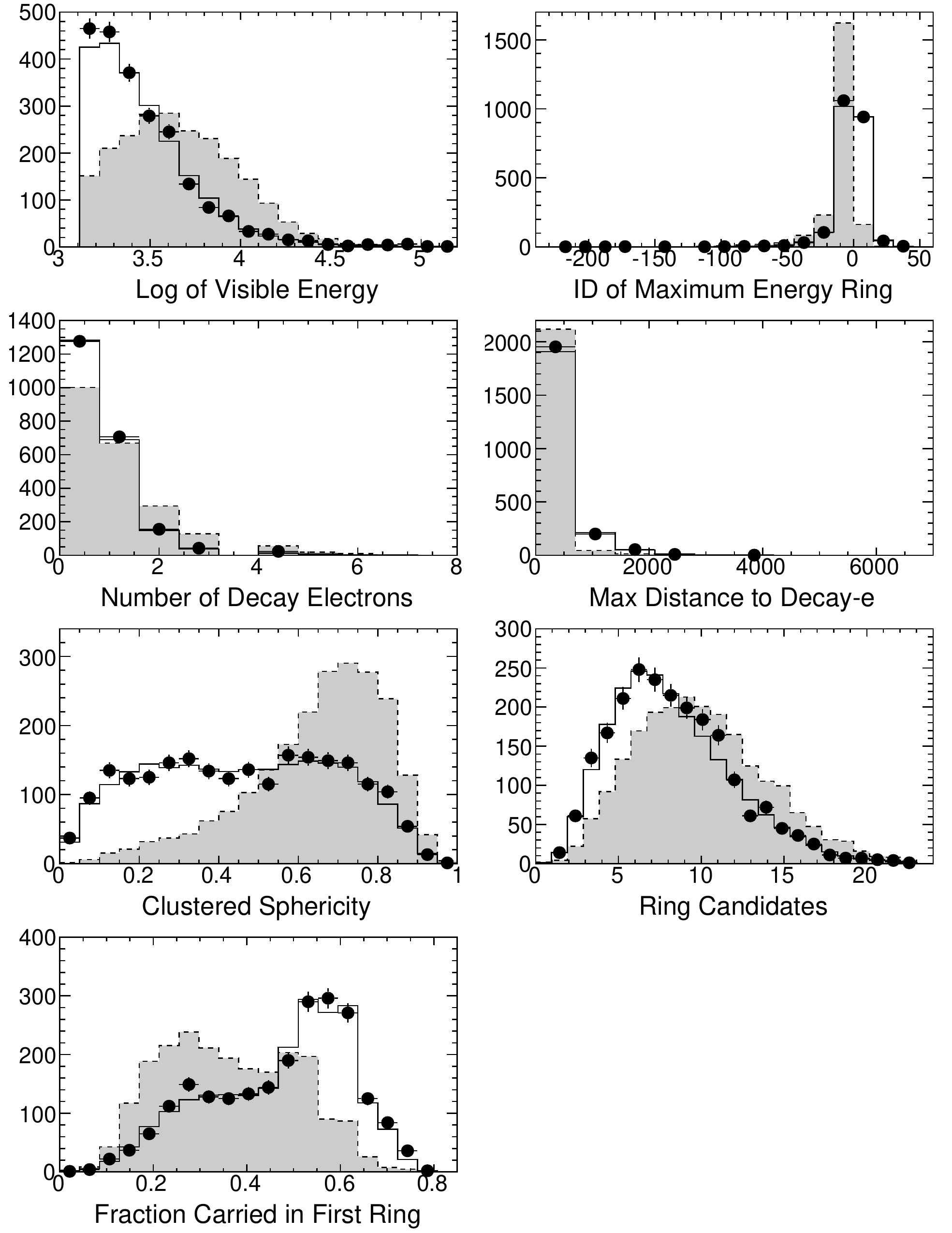}
  \caption{A figure showing the distributions of the seven input
    variables to the NN.  The downward going data (points) are
    overlaid with the downward going atmospheric MC (solid line). Also
    shown is the tau signal MC (shaded).  The variables in each plot
    are explained in the text. The tau signal is normalized for equal
    statistics.  The output of this NN is shown as figure 1 in the
    manuscript.}
  \label{fig:variables}
\end{figure}

\end{document}